\documentstyle[12pt]{article}
\input tables
\def\ymin{y_{\rm min}}

\begin{document}

\begin{titlepage}
\begin{flushright}
DTP/94/80\\
\end{flushright}
\vspace{2cm}
\begin{center}
{\Large\bf The Energy-Energy Correlation Function Revisited}\\
\vspace{1cm}
{\large
E.~W.~N.~ Glover and M.~R.~Sutton}\\
\vspace{0.5cm}
{\it
Physics Department, University of Durham,\\ Durham DH1~3LE, England} \\
\vspace{0.5cm}
\today
\vspace{0.5cm}
\end{center}
\begin{abstract}
The ${\cal O}(\alpha_s^2)$ coefficient of the energy-energy
correlation function (EEC) has been calculated by four groups with
differing results.  This discrepancy has lead to some confusion over
how to measure the strong coupling constant using the EEC and the
asymmetry of the energy-energy correlation function (AEEC) in
electron-positron annihilation at the $Z$ resonance.  For example, SLD
average the four values of $\alpha_s$ extracted from each of the
different calculations.  To resolve this situation, we present a new
calculation of this coefficient using three separate numerical
techniques to cancel the infrared poles.  All three methods agree with
each other and confirm the results of Kunszt and Nason that form the
benchmark for other ${\cal O}(\alpha_s^2)$ quantities.  As a
consequence, the central values and theoretical errors of the strong
coupling constant derived by SLD from the EEC and AEEC are altered.
Using the SLD data, we find, $\alpha_s^{EEC}(M_Z^2) =
0.125^{+0.002}_{-0.003}~({\rm exp.})
\pm 0.012 ~({\rm theory})$ and
$\alpha_s^{AEEC}(M_Z^2) = 0.114\pm 0.005~({\rm exp.})
\pm 0.004 ~({\rm theory})$.

\end{abstract}
\end{titlepage}

The energy-energy correlation function $\Sigma_{EEC}$ has recently
been used to measure the strong coupling constant, $\alpha_s$, in
$e^+e^-$ annihilation on the $Z$ resonance\footnote{A significant
amount of data has also been collected at lower energies, see for
example
\cite{SW} and references therein.} \cite{delphi,opal,opalnew,sld}.
It is defined in terms of the angle $\chi_{ij}$
between two particles $i$ and $j$ such that,
\begin{equation}
\frac{1}{\sigma}
\frac{d\Sigma_{EEC}(\chi)}{d\cos\chi} =
\frac{1}{\Delta\cos\chi~N_{events}}
\sum_{N_{events}}
\sum_{ij} \frac{E_iE_j}{E^2},
\end{equation}
where $E_i$ and $E_j$ are the energies of the particles and $E$ the
total energy in the event, $E = \sqrt{s}$.  The sum runs over all
pairs $ij$ lying within a bin in $\cos\chi$ of width $\Delta\cos\chi$
so that $\cos\chi-\Delta\cos\chi/2 < \cos\chi_{ij} <
\cos\chi+\Delta\cos\chi/2$.
For $i\neq j$, each pair enters twice in the sum so the integral of
the distribution is normalised to one when integrated over the whole
range of $\chi$.

The energy-energy correlation function can be
 described by a perturbative series,
\begin{equation}
\frac{1}{\sigma}
\frac{d\Sigma_{EEC}(\chi)}{d\cos\chi} =
\left(\frac{\alpha_s(\mu)}{2\pi}\right)
A(\chi)+  \left(\frac{\alpha_s(\mu)}{2\pi}\right)^2
\left(2\pi b_0 \log\left(\frac{\mu^2}{s}\right)A(\chi)
+ B(\chi)\right ) + {\cal O}(\alpha_s^3),
\end{equation}
where $b_0 = (11N-2n_f)/12\pi$ and $\sigma$ is the first order hadronic
cross section,
$$
\sigma = \sigma_0 \left (1 + \frac{\alpha_s(\mu)}{\pi}\right ),
$$
with $\sigma_0$ being the Born cross section for $e^+e^- \to $
hadrons.  More often, the energy-energy correlation function is
described relative to the Born cross section,
\begin{equation}
\frac{1}{\sigma_0}
\frac{d\Sigma_{EEC}(\chi)}{d\cos\chi}
= \left(\frac{\alpha_s(\mu)}{2\pi}\right)
\bar A(\chi)
+  \left(\frac{\alpha_s(\mu)}{2\pi}\right)^2 \left(2\pi b_0
\log\left(\frac{\mu^2}{s}\right)\bar A(\chi) + \bar B(\chi)\right )
+ {\cal O}(\alpha_s^3),
\end{equation}
where $\bar A = A$, $\bar B = B+2A$.  The lowest order contribution to
the energy-energy correlation function for $0^\circ < \chi <
180^\circ$ is obtained from the $Z,~\gamma^* \to q\bar qg$ process,
where $\chi$ can be the angle between any of the three partons.  An
analytic form for the leading order coefficient $A$ has been obtained
by Basham et al \cite{BBEL},
\begin{equation}
\bar A(\chi) = A(\chi) = (1+\omega)^3 \frac{1+3\omega}{3\omega}
\left[
(2-6\omega^2)\log\left(1+\frac{1}{\omega}\right) + 6\omega-3\right],
\end{equation}
where $\omega = \cot^2(\chi/2)$.  At next-to-leading order, the
relevant processes are $Z, ~\gamma^* \to q\bar q g$ at one-loop and
$Z,~\gamma^* \to q\bar q q\bar q,~~ q\bar q gg$ at tree level.  The
matrix elements for these processes were first computed by Ellis, Ross
and Terrano (ERT) \cite{ERT} in the $\overline{MS}$ scheme and have
formed the basis for all subsequent estimates of the second order
coefficient $\bar B$.  However, $\bar B$ has been computed several
times with differing results, first by Ali and Barreiro (AB)
\cite{AB1,AB2} and by Richards, Stirling and Ellis (RSE)
\cite{RSE1,RSE2} and more recently by Falck and Kramer (FK) \cite{FK}
and Kunszt and Nason (KN) \cite{KN}.  Despite this disagreement, the
radiative corrections are known to be relatively large while
hadronisation corrections are also significant. Therefore, to extract
the strong coupling constant, it is more usual to define the
asymmetry,
\begin{equation}
\frac{1}{\sigma}
\frac{d\Sigma_{AEEC}(\chi)}{d\cos\chi} =
\frac{1}{\sigma}
\frac{d\Sigma_{EEC}(180^\circ-\chi)}{d\cos\chi}-
\frac{1}{\sigma}
\frac{d\Sigma_{EEC}(\chi)}{d\cos\chi},
\end{equation}
where the corrections are smaller and the intra-jet region at $\chi
\sim 0^\circ $ and the back-to-back two jet region at $\chi \sim
180^\circ$ are suppressed.

Nevertheless, the disagreement of the theoretical calculations has
caused confusion in the extraction of $\alpha_s$ using the ${\cal
O}(\alpha_s^2)$ coefficient from both the EEC and AEEC.  For example,
OPAL use the Kunszt-Nason calculation to extract a central value for
$\alpha_s$ from the AEEC with a theoretical error on
$\Lambda_{\overline{MS}}$ of +55~Mev/-10~MeV to encompass the range of
predictions
\cite{opal}.
On the other hand, SLD average the values of $\alpha_s$ obtained from
the four theoretical calculations and increase the theoretical error
accordingly \cite{sld}.  They find,
\begin{eqnarray}
\alpha^{EEC}_s(M_Z^2) & = & 0.127^{+0.002}_{-0.003}~({\rm exp.})
\pm 0.013 ~({\rm theory}),\nonumber \\
\alpha^{AEEC}_s(M_Z^2) & = & 0.116\pm 0.005~({\rm exp.})
\pm 0.006 ~({\rm theory}),\nonumber
\end{eqnarray}
where the theoretical error includes the uncertainty from
hadronisation ($\pm 0.002$ for EEC and $^{+0.003}_{-0.002}$ for AEEC),
renormalisation scale ($\pm 0.011$ for EEC and $\pm 0.003$ for AEEC)
and an error of $\pm 0.006$ for EEC and $\pm 0.004$ for the AEEC
purely from the uncertainty over which ${\cal O}(\alpha_s^2)$
coefficient is correct \cite{sld}.

In this letter, we attempt to resolve the theoretical disagreement.
We first review the discrepancies amongst the existing calculations of
$\bar B$.  We then present a new calculation of the energy-energy
correlation function at ${\cal O}(\alpha_s^2)$ which is in complete
agreement with the results of Kunszt and Nason.

In order to compute next-to-leading order quantities in perturbation
theory, it is necessary to combine the contribution from $n$-parton
one-loop Feynman diagrams with the $n+1$-parton bremstrahlung process.
The virtual matrix elements are divergent and contain both infrared
and ultraviolet singularities.  The ultraviolet poles are removed by
renormalisation, however the soft and collinear infrared poles are
only cancelled when the virtual graphs are combined with the
bremstrahlung process.  Although the cancellation of infrared poles
can be done analytically for simple processes, for complicated
processes like this, it is necessary to resort to numerical
techniques.  Since the theoretical calculations of $\bar B$ are all
based on the same ERT matrix elements, the discrepancies amongst the
theoretical calculations appear to be rooted in the numerical
implementation.

The numerical problem has been nicely formulated by Kunszt and Soper
\cite{KS} by means of a simple example integral,
\begin{equation}
{\cal I} = \lim_{\epsilon \to 0} \left \lbrace
\int^1_0 \frac{dx}{x} x^\epsilon F(x) -
\frac{1}{\epsilon} F(0)\right \rbrace,
\end{equation}
where $F(x)$ is a known but complicated function representing the
$n+1$-parton bremstrahlung matrix elements.  Here $x$ represents the
angle between two partons or the energy of a gluon and the integral
over $x$ represents the additional phase space of the extra parton.
As $x\to 0$, the integrand is regularised by the $x^\epsilon$ factor
as in dimensional regularisation, however, the first term is still
divergent as $\epsilon \to 0$.  This divergence is cancelled by the
second term - the $n$-parton one-loop contribution - so that the
integral is finite.  A variety of methods to compute ${\cal I}$ have
been developed.

The original method used by ERT is also known as the subtraction
method.  Here, a divergent term is subtracted from the first term and
added to the second,
\begin{eqnarray}
{\cal I} &= & \lim_{\epsilon \to 0} \left \lbrace
\int^1_0 \frac{dx}{x} x^\epsilon (F(x)-F(0))
+F(0) \int^1_0 \frac{dx}{x} x^\epsilon
- \frac{1}{\epsilon} F(0)\right \rbrace \nonumber \\
    &= & \int^1_0 \frac{dx}{x}  \left (F(x)-F(0)\right),
\end{eqnarray}
so that the integral is manifestly finite.  This method has the
advantages of requiring no extra theoretical cutoffs and making no
approximations.  A disadvantage is that it does require some
additional analytic effort to explicitly extract the poles from the
analogue of $\int^1_0 \frac{dx}{x} x^\epsilon$.  For ${\cal
O}(\alpha_s^2)$ quantities in electron-positron collisions, this was
performed by Ellis, Ross and Terrano \cite{ERT}.  However, this
analytic integration has to be carried out from scratch for each
process under investigation and may even require a knowledge of the
experimental jet algorithm \cite{KS}.

An alternate approach known as the phase space slicing method has been
widely used - see ref.~\cite{GG} and refences theirin.  The
integration region is divided into two parts, $0 < x < \delta$ and
$\delta < x < 1$.  In the first region, the function $F(x)$ can be
approximated by $F(0)$ provided the arbitrary cutoff $\delta \ll 1$,
\begin{eqnarray}
{\cal I} &\sim & \lim_{\epsilon \to 0} \left \lbrace
\int^1_\delta \frac{dx}{x} x^\epsilon F(x)
+F(0) \int^\delta_0 \frac{dx}{x} x^\epsilon
- \frac{1}{\epsilon} F(0)\right \rbrace \nonumber \\
    &\sim &
\int^1_\delta \frac{dx}{x}   F(x)
+F(0) \ln(\delta).
\end{eqnarray}
This method is extremely portable \cite{GG} since the soft and
collinear approximations of the matrix elements and phase space are
universal.  This makes it easy to apply to a wide variety of
physically interesting processes.  However, the disadvantage is the
presence of the arbitrary cutoff $\delta$.  The integral should not
depend on $\delta$, and the $\delta$ dependence of the two terms in
Eq.~8 should cancel.  Since the approximations are reliable only when
$\delta$ is small, this can give rise to numerical problems.

Finally, a third method is a hybrid of the two previous techniques.
To preserve the portability of the phase space slicing method, we add
and subtract only the universal soft/collinear approximations for $x <
\delta$,
\begin{eqnarray}
{\cal I} &\sim & \lim_{\epsilon \to 0} \left \lbrace
\int^1_0 \frac{dx}{x} x^\epsilon F(x)
-F(0) \int^\delta_0 \frac{dx}{x} x^\epsilon
+F(0) \int^\delta_0 \frac{dx}{x} x^\epsilon
- \frac{1}{\epsilon} F(0)\right \rbrace \nonumber \\
    &\sim &
\int^1_0 \frac{dx}{x}  \left( F(x)-F(0) \theta(\delta-x) \right)
+F(0) \ln(\delta).
\end{eqnarray}
A cancellation between the terms still occurs, however only the phase
space is approximated, so that this method is valid at larger values
of $\delta$.  The difference between the latter two approaches is,
\begin{equation}
\int^\delta_0 \frac{dx}{x}  \left( F(x)-F(0) \right),
\end{equation}
which clearly tends to zero as $\delta \to 0$.  Therefore, provided
$\delta$ is chosen small enough, all three methods should give
equivalent results.

We will present results using all three methods, however, we first
turn briefly to a discussion of the previous calculations, each of
which uses a different notation for the first and second order
coefficients $\bar A$ and $\bar B$ as shown in Table~1.  We follow the
event shape description and focus on the perfect resolution limit
($\delta \to 0$). In other words, no jet definition is applied to the
partons before computing the energy-energy correlation function.

Both AB and RSE use the subtraction method to compute $\bar B$.  AB
perform the five dimensional integral over the four parton phase space
numerically, while RSE relate the angle between partons $\chi$ to the
invariants and are left with a four dimensional integral.  Their
results for the energy-energy correlation at $\chi \sim 90^\circ$ are
shown in Table~2.  Because one of the integrations has been removed,
RSE have significantly smaller errors.  Nevertheless, within the
errors, both agree.  This is in contrast to the claim in \cite{AB2}
where $B/A$ of RSE is compared with $\bar B/\bar A$ of
AB\footnote{Table~2 of \cite{AB2} claims to show the ratio $R^{corr}
\sim B/ A $ however inspection of the raw numbers in Table~1
indicates that actually $R^{corr} \sim \bar B/\bar A$ is quoted.  As
a result the comparison of AB and RSE in Fig.~4 of \cite{AB2} is
flawed.  Unfortunately this claim has propagated through the
literature \cite{FK,KN}.}.

\begin{table}[t]
\begintable
             | $\bar A$             | $\bar B$             \crthick
AB$^{(1)}$   | $2F$                 | $4G$                 \cr
RSE$^{(2)}$  | $2g^{(1)}$           | $4g^{(2)}$           \cr
FK$^{(3)}$   | $2C$                 | $4D$                 \cr
KN$^{(4)}$   | $A_{EEC}/\sin^2\chi$ | $B_{EEC}/\sin^2\chi$ \endtable
\caption[]{The definitions of the LO and NLO coefficients of the
different calculations;\\
 (1) Eq. 1.6 of ref.~\cite{AB2},
(2) eq. 1.4 of ref.~\cite{RSE2}, (3) eq. 1.2 of ref.~\cite{FK},
(4) eq. 4.24 of ref.~\cite{KN}. }
\end{table}

Falck and Kramer \cite{FK} present two results for the perfect
resolution limit.  The first is based on a phase space slicing
approach with a theoretical cutoff such that $y_{ij} = (p_i+p_j)^2/s >
\ymin = 10^{-4}$.  For such small values of $\ymin$, it is assumed
that the limit $\ymin \to 0$ can be considered to have been reached -
an assumption supported by earlier studies \cite{AB1}.  This result is
somewhat larger than AB and RSE, but the numerical errors also appear
to be larger.  As a check, FK quote a second much smaller result based
on the subtraction method which is in rough agreement with RSE (and
hence AB).  However, in conclusion, FK ascribe these differences to
the presence of the cutoff $\ymin$ and the phase space slicing method
used to compute $\bar B/\bar A$.

\begin{table}[t]
\begintable
            | $\bar A(\chi \sim 90^\circ)$ | $\bar B(\chi \sim 90^\circ)$
 | $\bar B/\bar A$\crthick
AB$^{(1)}$  | 2.434 | $42.68 \pm 1.94$ | $17.53 \pm 0.79$      \cr
RSE$^{(2)}$ | 2.426 | $41.52 \pm 0.16$ | $17.1  \pm 0.06 $     \cr
FKa$^{(3)}$ | $--$  | $--$             | $25.1  \pm ?$         \cr
FKb$^{(4)}$ | $--$  | $--$             | $17.5  \pm ?$         \cr
KN$^{(5)}$  | 2.43  | $51.25\pm 2.67$  | $21.1  \pm 1.1$       \cr
GSa$^{(6)}$        | 2.43  | $52.39\pm 0.83$  | $21.6  \pm 0.34$
   \cr
GSb$^{(7)}$         | 2.43  | $51.15\pm 0.68$  | $21.05 \pm  0.28$
    \cr
GSc$^{(8)}$         | 2.43  | $52.29\pm 2.08$  | $21.52 \pm 0.86$
   \endtable
\caption[]{The NLO to LO coefficients and the ratio
$\bar B/\bar A$ for the different calculations;
(1) Table 1 of ref.~\cite{AB2},
(2) Table 3 of ref.~\cite{RSE2},
(3) Fig.~ 3 of ref.~\cite{FK},
(4) Fig.~ 6 of ref.~\cite{FK},
(5) Tables 2 and 3 of ref.~\cite{KN},
(6) The subtraction method,
(7) The phase space slicing scheme with $\ymin=10^{-5}$,
(8) The hybrid subtraction scheme with $\ymin=10^{-5}$.}
\end{table}

Finally, the benchmark calculation of ${\cal O}(\alpha_s^2)$
quantities at LEP energies is that of Kunszt and Nason.  By using a
sophisticated phase space mapping, they have reorganised the ERT
matrix elements to give numerically stable results for all of the
event shape and three jet quantities measured at LEP.  Many of their
predictions for other quantities have been checked
\cite{glo} and it would be rather surprising for a single distribution
to be in error.  However, despite claiming to agree with FK, the KN
result at $\chi \sim 90^\circ$ lies between the other calculations.

We have recalculated the ${\cal O}(\alpha_s^2)$ coefficient using all
three numerical techniques described earlier.  First, we have recoded
the ERT matrix elements precisely as described in \cite{ERT}.
However, rather than weight each event by the value of the $C$
parameter \cite{ERT}, we have weighted by the energy-energy
correlation function,
$$
\Sigma^{(3)} = \Sigma_{ij, k\neq i,j}
\frac{2E_iE_j}{s} ~\delta\left(\cos\chi -
\left(\frac{y_{ik}y_{jk}-y_{ij}}{y_{ik}y_{jk}+y_{ij}}\right)\right),
$$
and,
$$
\Sigma^{(4)} = \Sigma_{ij, k\neq i,j, l \neq i,j,k} \frac{2E_iE_j}{s}
{}~\delta\left(\cos\chi -
\left(\frac{y_{ikl}y_{jkl}-y_{kl}-y_{ij}}
{y_{ikl}y_{jkl}-y_{kl}+y_{ij}}\right)\right),
$$
for three and four parton final states respectively.

\begin{figure}\vspace{8cm}
\includegraphics{fig1.ps}
\caption[]{The ratio $\bar B/\bar A$ at $\chi \sim 90^\circ$
as a function of $\ymin$ for the phase space slicing (GSb) and hybrid
subtraction (GSc) schemes.  The $\ymin$ independent values obtained
from the subtraction method (GSa) and from ref.~\cite{KN} are also
shown.}
\end{figure}

Second, we have constructed a completely independent program along the
lines of \cite{GG} using squared matrix elements rather than helicity
amplitudes.  Either phase space slicing or the hybrid subtraction
scheme may be selected.  As described earlier, these methods rely on
an unphysical cut to isolate the divergences.  Fig.~1 shows the ratio
$\bar B/\bar A$ at $\chi \sim 90^\circ$ as a function of this cut,
$\ymin$.  At large $\ymin$, the predictions using the phase space
slicing method varies rapidly with $\ymin$.  This is because the
approximations used to perform the analytic integrations are
inaccurate.  However, despite the increasing numerical errors, we see
that for $\ymin < 10^{-4}$, the variation with $\ymin$ is small.  A
reasonable approximation to the $\ymin \to 0$ limit is therefore
$\ymin = 10^{-5}$.  In the hybrid subtraction scheme, $\bar B/\bar A$
also varies rapidly when $\ymin$ is large,
but the zero resolution limit is also
approximated by $\ymin = 10^{-5}$.

Table~2 shows the value of $\bar B/\bar A$ at $\chi \sim 90^\circ$ for
these three methods.  Within errors, all three agree with each other
and with the result of Kunszt and Nason.  In principle, the
calculation of Falck and Kramer (FKa) should coincide with those of
the phase space slicing method for $\ymin= 10^{-4}$, and the results
of AB and RSE should agree with the subtraction calculation, GSa,
however, we see that this is not the case.

\begin{figure}\vspace{8cm}
\includegraphics{fig2.ps}
\caption[]{The ratio of KN \cite{KN}
and our results using the phase space slicing method with
$\ymin = 10^{-5}$
(GSb) over the whole range of $\cos\chi$.}
\end{figure}

So far we have concentrated on a single value of $\chi$.  Fig.~2 shows
the ratio of the Kunszt-Nason calculation and our phase space slicing
results with $\ymin = 10^{-5}$ over the whole range of $\cos\chi$.  We
see that the two calculations are in good agreement with errors of
less than ${\cal O}(5\%)$.  Fitting a constant to this data
yields a ratio of 1.0068 with a $\chi^2$/d.o.f of 0.98.  The other
numerical methods give similar results.

We therefore conclude that the KN calculation is indeed correct and
that the other predictions (AB, RSE and FK) seem to be deficient in
some way.  This has a direct impact on the measurements of the strong
coupling constant made at LEP and SLC.  For example, by eliminating
this source of theoretical error, the values of $\alpha_s$ extracted
from the EEC and AEEC quoted earlier become\footnote{See Table 1 of
\cite{sld}},
\begin{eqnarray}
\alpha^{EEC}_s(M_Z^2) & = & 0.125^{+0.002}_{-0.003}~({\rm exp.})
\pm 0.012 ~({\rm theory}),\nonumber \\
\alpha^{AEEC}_s(M_Z^2) & = & 0.114\pm 0.005~({\rm exp.})
\pm 0.004 ~({\rm theory}),\nonumber
\end{eqnarray}
where, in addition to a reduction in the quoted theoretical error, the
central values have also changed.

Finally, the ${\cal O}(\alpha_s^2)$ coefficients have also provided
an input into QCD calculations where logarithms of the form $\log(1/y)$
where $y = \frac{1+\cos\chi}{2}$ have been resummed \cite{catani}.
It is worth noting that the coefficients of the logarithmic terms
computed using resummation techniques have been shown to
agree with the coefficients extracted from the numerical results of
Kunszt and Nason  and not with those of AB, RSE and FK \cite{turnock}.
This provides additional confirmation of the results presented here and
of the validity of the resummation method.

SLD have obtained a value of $\alpha_s$ using such resummed
calculations for the EEC,
again averaging over the four calculations of $\bar B$,
$$
\alpha^{EEC}_s(M_Z^2) = 0.130^{+0.003}_{-0.004}~({\rm exp.})
\pm 0.007 ~({\rm theory}) ~~~~~({\rm resummed}).
$$
Using the KN
calculation alone, this becomes \cite{sld},
$$
\alpha^{EEC}_s(M_Z^2) = 0.129^{+0.003}_{-0.004}~({\rm exp.})
\pm 0.005 ~({\rm theory})~~~~~({\rm resummed}),
$$
which, within errors, is consistent with the measurements from the
EEC and AEEC using the ${\cal O}(\alpha_s^2)$ calculations.

In conclusion, motivated by an apparent disagreement amongst
theoretical calculations of the ${\cal O}(\alpha_s^2)$ coefficient of
the energy-energy correlation function, we have recomputed it using
three different numerical techniques.  With all three methods, we
reproduce the results of Kunszt and Nason which have formed the
benchmark for extracting a value for the strong coupling constant from
LEP and SLC data.\\

\noindent{ \Large \bf Acknowledgements}

EWNG thanks James Stirling for enlightening discussions and Ramon
Munoz-Tapia for critically reading the manuscript.

\end{document}